\documentstyle[prl,aps,epsfig]{revtex} \draft
\begin{document}

\twocolumn[
\hsize\textwidth\columnwidth\hsize\csname@twocolumnfalse\endcsname

\title{EVIDENCE\ FOR\ DISTINCT\ POLYMER\ CHAIN\
ORIENTATIONS\ IN\ KC$_{60}$\ AND\ RbC$_{60}$}

\author{P.\ Launois, R.\ Moret } \address{Laboratoire de
Physique des Solides (URA CNRS 2), Universit{\'e}
Paris--Sud, B{\^a}t. 510, 91405 Orsay CEDEX, FRANCE}
\author{J.\ Hone, A.\ Zettl} \address{Department of Physics,
University of California at Berkeley,\\and Materials
Sciences Division, Lawrence Berkeley National Laboratory,
Berkeley, California 94720, USA\\} 
\date{september 1998: to be published in Phys. Rev. Lett.}
\maketitle

\begin{abstract} The KC$_{60}$ and RbC$_{60}$ polymer phases
exhibit contrasting electronic properties while powder
diffraction studies have revealed no definite structural
difference. We have performed  single crystal X-ray
diffraction and diffuse scattering studies of these
compounds. It is found that KC$_{60}$ and RbC$_{60}$ possess
different chain orientations about their axes, which are
described by distinct space groups Pmnn and I2/m,
respectively. Such a structural difference will be of  great
importance to a complete understanding of the physical
properties. \end{abstract}

\vfill  \twocolumn \vskip.5pc ]
\narrowtext

The recently discovered alkali-fullerides
AC$_{60}$\cite{dec} (A=K, Rb, Cs) exhibit a phase transition
from a high temperature cubic phase \cite {trans} to an
orthorhombic one in which the molecules form one-dimensional
polymer chains\cite{chauvet,chainpekk,stephens,fox}, at
about $350K$. Despite extensive studies, the physical
properties of the latter phase are still  the subject of
controversy. They have been investigated by
ESR\cite{chauvet,bommeli,auban,janossy}, $\mu $SR\cite{muon}
, NMR\cite{auban,alloul,brouet,rachdi}, and optical and
electrical conductivity measurements\cite{bommeli,hone}.
RbC$_{60}$ and CsC$_{60}$ possess a magnetic transition
towards an insulating phase below $\sim 50K$, whereas KC$
_{60}$ does not. The exact nature of the magnetic low
temperature phase is not yet understood; several scenarios
for quasi-one-dimensional \cite
{chauvet,chainpekk,bommeli,janossy,brouet} or three
dimensional \cite {auban,erwin,kuzmany} magnetic ordering
are being debated. The different behavior of KC$_{60}$
relative to RbC$_{60}$ and CsC$_{60}$ is also not
understood. Recent theoretical calculations show that the
chain orientations influence the dimensionality of the
electronic properties\cite {tanaka}. However, powder
diffraction studies\cite{stephens} revealed no definite
structural difference between KC$_{60}$ and RbC$_{60}$.
Accordingly, a better knowledge of the details of the
AC$_{60}$ structures, especially the chain orientations, is
needed. We present the first \emph{\ }single crystal\emph{ \
}diffraction study of KC$_{60}$ and RbC$_{60}$, and we show
that KC$_{60}$ and RbC$_{60}$ possess different relative
chain orientations.

The main structural results\cite{chauvet,stephens,fox}
obtained for KC$_{60}$ and RbC$_{60}$ are summarized in the
following.\ The unit cell is orthorhombic, with parameters
$a$, $b$ and $c$ equal to $9.11,9.95,14.32$ \AA\ and to
$9.14,10.11,14.23$\AA\ for KC$_{60}$ and RbC$_{60}$,
respectively \cite{stephens}. C$_{60}$ molecules are
centered at $(0,0,0)$ and $(1/2,1/2,1/2)$ positions, and
alkali ions at $(0,0,1/2)$ and $ (1/2,1/2,0)$.
Polymerization occurs via [2+2] cycloaddition along the
shortest parameter $\bf{a}$ (fig.1(a)). The most plausible
orthorhombic space groups compatible with the molecular
symmetry are Immm 

\begin{figure}[ht]
\centerline{\epsfig{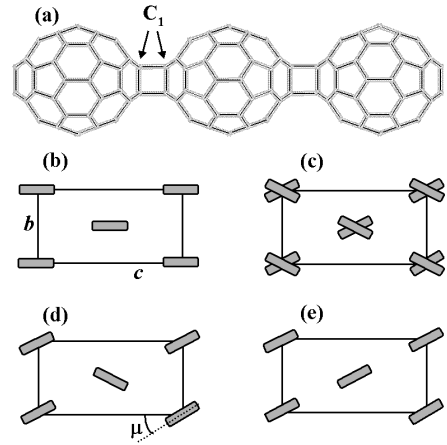}}
\vskip 0.2cm
\caption{(a) Linear polymer chain formed
by [2+2] cycloaddition. Schematic drawing of chain
orientations for (b) ordered Immm, (c) disordered Immm, (d)
Pmnn, (e) I2/m. The shaded bars indicate the orientation of
the polymer chains, they represent the projection of the
cycloaddition planes defined by the C$ _{1}$ atoms onto the
crystallographic $(\bf{b},\bf{c})$ plane.}
\end{figure}

and Pmnn.\ The orientation of a C$_{60}$
chain about its axis $\bf{a}$ can be characterized by the
angle $\mu $ of the planes of cycloaddition with $\bf{c}$.
For the body-centered space group Immm, two configurations
must be considered: an ordered one, with $\mu =0$ or
$90{^{\circ }} $ (fig.1(b)) where chains passing through the
origin and the center of the unit cell have the same
orientation, and a disordered one if $\mu \neq 0$ or $90{
{}^{\circ }}$: the mirror planes perpendicular to $b$ and
$c$ constrain the chains to take orientations $\mu $ or
$-\mu $, with equal probabilities (fig.1(c)). Such a
disorder would give rise to diffuse planes perpendicular to
$\bf{a}^{*}$ in reciprocal space.\ The Pmnn structure has
glide planes, so that if the orientation of the chain
passing through the unit cell origin is $\mu $, the
orientation of that passing through its center is  $-\mu $
(fig.1(d)).\ In diffraction, Immm can be distinguished from
Pmnn by the extinction of the reflections for $h+k+l$ odd.
From Rietveld refinements, Stephens et al.\cite{stephens}
found $\mu =45\pm 5{ {}^{\circ }}$ for both KC$_{60}$ and
RbC$_{60}$ samples, but they could not discriminate between
Immm and Pmnn. A pair distribution function analysis
performed by Fox et al.\cite{fox} indicated a possible
orientational chain disorder.

For the present study, KC$_{60}$ and RbC$_{60}$ crystals
(typical size: $10^{-2}mm^{3}$) were prepared by
stoichiometric doping of C$_{60}$ single crystals at $400{
^{\circ }}C$; the polymer phase was obtained by subsequent
slow cooling (the detailed procedure is described in
ref.\cite{hone}). The samples were first characterized by
X-ray powder diffraction\cite{hone}(b) and electron beam
analysis\cite{chopra}; their electrical transport properties
were reported in ref.\cite{hone}. The sample crystallinity
suffered from the insertion and polymerization process, but
it was still acceptable for our diffraction studies (mosaic
spread $\approx 2{{}^{\circ }}$, full-width at half-maximum,
for KC$_{60}$ and $\approx 2.5{{}^{\circ }}$ for
RbC$_{60}$). 

The single-crystal X-ray experiments combined photographic
(precession and fixed-crystal) and diffractometer
techniques. The precession method, which gives undistorted
sections of the reciprocal space, enabled us to check the
body-centered extinctions mentioned above and to study the
domain structure of the crystal. The latter information
cannot be accessed by powder diffraction techniques. The
fixed-crystal fixed-film monochromatic technique was
employed to detect the X-ray diffuse scattering possibly due
to orientational disorder of the C$_{60}$ chains. This
technique, where the crystal is under vacuum, is
particularly efficient as it maximizes the signal/noise
ratio. Finally, the diffractometer technique enabled us to
make quantitative measurements of the Bragg peak intensities
used for structure refinements.

Precession photographs (CuK$\alpha $ radiation) have been
taken on different crystals to ascertain general results.
Complex precession patterns similar to those presented for
pressure polymerized C$ _{60}$ in ref.\cite{moret} were
obtained.\ They show the coexistence of orientational
variants due to the cubic-orthorhombic symmetry lowering.\
These variants are related by the lost symmetry operations.
We have determined the orientational relationships between
the variants, which gives information regarding the
structural polymerization mechanism. In KC$_{60}$, the
polymerization involves the sliding of dense $(111)_{c}$
cubic planes whose orientation is preserved, as in pressure
polymerized C$_{60}$\cite{moret}. The situation is somewhat
different for RbC$_{60}$ because the orientation of
$(111)_{c}$ planes is not preserved. The structural
polymerization mechanism observed for pressure polymerized
C$_{60}$ and for KC$_{60}$ does not apply to RbC$_{60}$
possibly due to steric constraints imposed by the larger Rb
ions. The symmetry elements preserved by the
cubic-orthorhombic transformation are (i) in KC$ _{60}$ the
$\left[ 110\right] _{c}=\bf{b}$ 2-fold axis and (ii) in
RbC$_{60}$ the  $\left[ 1\overline{1}0\right] _{c}=\bf{a}$
and the $\left[ 110\right] _{c}=\bf{b}$ 2-fold axes. This
transformation generates $12$ variants for KC$_{60}$ and
only $6$ for RbC$ _{60}$. A careful analysis of the
precession patterns reveals that KC$_{60}$ presents a
primitive (P) lattice while RbC$_{60}$ is body-centered (I,
absence of $h+k+l=2n+1$ reflections).

Experiments using the fixed-crystal fixed-film method were
performed for KC$_{60}$ and RbC$_{60}$. They revealed no
diffuse scattering.  Calculations showed that the diffuse
scattering intensity expected for chain disorder should have
the same order of magnitude as that produced by the rotating
molecules in pure C$_{60}$, which can be easily detected by
the fixed-crystal fixed-film method. We thus conclude that
the polymer chains are ordered in both KC$_{60}$ and
RbC$_{60}$.

The Bragg peak intensity measurements on KC$_{60}$ and
RbC$_{60}$ were performed on a three-circle diffractometer,
using CuK$\alpha $ radiation. AC$_{60} $ crystals often
present $\left\{ 111\right\} _{c}$ twins originating from
the parent C$_{60}$ crystals and we selected samples with
negligible twin volumes ($\leq 1\%$). The unit cell
parameters are found equal to those in ref. \cite{stephens}
within experimental accuracy. In order to refine the
structure, we had to measure diffraction peaks from a single
domain. This task required a selection procedure to exclude
overlapping reflections. First we computed the Bragg peak
positions for all variants and selected isolated
reflections, then we scanned these reflections towards
neighboring ones (typically within $0.8$ \AA $^{-1}$), to
ensure the absence of contamination. The remaining Bragg
peaks were fitted using gaussian profiles, yielding peak
intensities $I_{0}\left( hkl\right) $. We obtained the
following data set of unique and isolated reflections: $177$
reflections, among which $107$ reflections with $h+k+l=2n+1$
and $70$\ reflections with $h+k+l=2n$, for KC$_{60}$, $82$
reflections with $h+k+l=2n$ for RbC$ _{60}$. Their
intensities are relatively weak: only $111$ peaks for
KC$_{60}$ ( $63$ peaks with $h+k+l=2n+1$ and $48$ peaks with
$h+k+l=2n$) and $39$ peaks for RbC$_{60}$ verify the
relation $I>\sigma $\cite{stout}. We also checked i) the
$h+k+l=2n+1$ extinctions for RbC$_{60}$, ii) the glide plane
extinctions in KC$_{60}$.

The structural analysis is based on the minimization of the
reliability factor \cite{stout}  \begin{equation} R=\sum
\left| \frac{\left| F_{obs}\right| }{\sum \left|
F_{obs}\right| }- \frac{\left| F_{calc}\right| }{\sum \left|
F_{calc}\right| }\right|  \label{Rdef} \end{equation} where
$F_{obs}$ and $F_{calc}$ are observed and calculated
structure factors. $\left| F_{obs}\left( h,k,l\right)
\right| $ is the square-root of the integrated intensity
given by $I=I_{0}\left( hkl\right)\cdot Q^{2}$ (the mosaic
broadening of the reflection is considered to be
proportional to $Q^{2}$), and corrected for polarization
effects. The limited intensity data lead us to restrict the
number of refined parameters to the chain orientation angle
$\mu $, an isotropic carbon Debye-Waller (DW) parameter
$U_{C}$ , anisotropic alkali DW parameters
$U_{11},U_{22},U_{33}$, and the molecular distortion; we
relaxed the C$_{1}$ atom positions only, which most affects
the reliability factor\cite{stephens}. We varied these
parameters simultaneously over broad ranges using a step by
step procedure which required reasonably small computing
times due to the limited data set. The set of parameter
values that gave the lowest R ($R_{\min }$) was retained. In
comparison with the usual least-squares refinement method,
this procedure enables us to test all possible combinations
of the parameters, but its drawback is that uncertainties
are not easily evaluated.  

The R values obtained by minimization over the molecular
distortion and over the DW factors are plotted versus
orientation angle $\mu$ in fig.2. For Pmnn KC$_{60}$, we
obtain $R_{\min }\simeq 0.16$ for $ U_{C}\simeq 0.01$\AA
$^{2}$, $U_{11},U_{22},U_{33}\simeq 0.24,0.02,0.3$\AA $
^{2}$, $d_{C_{1}-inter}\simeq d_{C_{1}-intra}\simeq 1.55$\AA
, and $\mu \simeq 51 {{}^{\circ }}$. The DW factor for
carbon is normal, while those of potassium are unusually
large; preliminary experiments as a function of temperature
indicate that they probably conceal alkali ion
displacements\cite {ulter}.

\begin{figure}[th]
\centerline{\epsfig{file=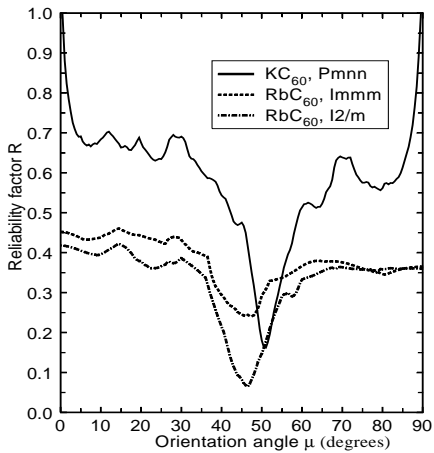,angle=0.0,height=8cm,width=7cm}}
\vskip 0.2cm
\caption{Reliability factor R calculated
for peaks satisfying $I>\sigma $, as a function of the chain
orientation $\mu $ (see the text).}
\end{figure}

For RbC$_{60}$ a refinement was
first attempted within the Immm space group, leading to
$R_{\min }\simeq 0.24$ for $\mu \simeq 48{{ {}^{\circ
}}}$(fig.2). However this implies, as indicated before, a $
+\mu /-\mu $ orientational chain disorder which is ruled out
by the absence of diffuse scattering. Aside from Immm, no
body-centered orthorhombic space group is convenient to
describe the orientational ordering of C$_{60}$ polymer
chains. We were thus forced to consider monoclinic
body-centered arrangements. The space group compatible with
the symmetry of C$_{60}$ chains is I2/m, with the chain axis
parallel to the 2-fold axis; the corresponding chain
orientations are depicted in fig.1(e). In this case, the
$(\bf{b},\bf{c})$ angle is not constrained to $90{{}^{\circ
}}$. However, we have not  detected a deviation of this
angle value from $90{{}^{\circ }}$ (within an estimated
experimental uncertainty of $0.5{{}^{\circ}}$). It may be
very weak because of the relatively homogeneous distribution
of the atoms on a chain around its axis (if  it were fully
homogeneous, the $(\bf{b},\bf{c})$ angle would be equal to
$90{{}^{\circ }}$). Within the I2/m hypothesis, the
reliability factor minimum $R_{\min }\simeq 0.06$
corresponds to: $U_{C}\simeq 0.01$\AA $^{2}$,
$U_{11},U_{33}\simeq 0.16,0.09$ \AA
$^{2}$\cite{absorb}($U_{22}$ cannot be determined because
all measured $h,k,l$ peaks have small $k$ values),
$d_{C_{1}-inter}\simeq 1.5$\AA, $ d_{C_{1}-intra}\simeq
1.6$\AA , and\emph{\ }$\mu \simeq 47{{}^{\circ }}$
(fig.2)\cite{comment}. This model is highly attractive
because the chain orientations are ordered (in agreement
with the absence of diffuse scattering) and it greatly
improves $R_{\min }$, as compared to Immm. The chain
orientations in KC$_{60}$ and RbC$_{60}$ ($\mu \simeq
51{{}^\circ}$ and $47{{}^\circ}$)  are in agreement with the
results of Stephens et al. ($45\pm 5
{{}^\circ}$)\cite{stephens}, and the molecular distortions
are found to be similar in KC$_{60}$ and
RbC$_{60}$\cite{fox}. However the different space groups
obtained for KC$_{60}$ and RbC$_{60}$, namely Pmnn and I2/m,
imply completely different relative orientations of the
chains.

\begin{figure}[th]
\centerline{\epsfig{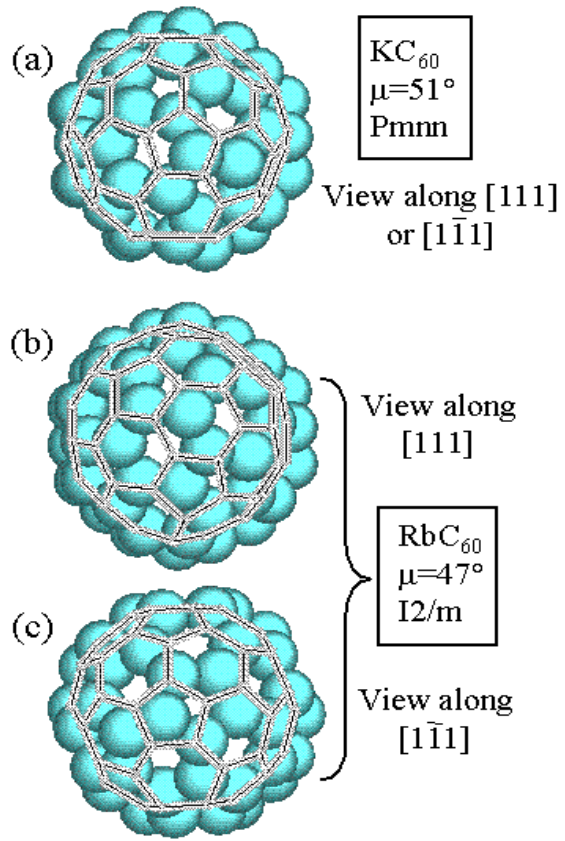}}
\vskip 0.2cm
\caption{Molecular environments viewed
along the axes $\left[111\right]$ and
$\left[1\overline{1}1\right]$ joining centers of nearest
neighbor molecules in KC$_{60}$ and in RbC$_{60}$. The C
atoms (spheres) of the more distant molecule are viewed
through a portion of the C-C bond skeleton of the nearer
one.} 
\end{figure}

It is interesting to compare the structural environments in
KC$_{60}$ and RbC$_{60}$. The first neighbor
C$_{60}$-C$_{60}$ interchain ($8.73$\AA) and intermolecular
($9.85$\AA ) distances are remarkably similar for both
compounds while the second neighbor distance (equal to b)
increases from $9.95$\AA\ (KC$_{60}$) to $10.11$\AA\
(RbC$_{60}$). The A-C$_{60}$ distance increases for the
first neighbors ($6.74$\AA\ for KC$_{60}$ and $6.81$\AA\ for
RbC$_{60}$) while it decreases slightly for the second
neighbors ($7.16$\AA\ for KC$_{60}$ and $7.12 $\AA\ for
RbC$_{60}$). The above distances depend on the unit cell
parameters only. The chain orientation comes into play for
the interatomic A-C or C-C environments. In both compounds
the alkali ions roughly face carbon hexagons from the first
neighbors C$_{60}$ (along $\left[ 110\right]$ and
$\left[1\overline{1}0\right]$) and 'single' C-C bonds from
their second neighbors (along
$\left[001\right]$)\cite{molec}. The influence of the
different values of $\mu$ ($\mu_{K} \simeq 51{{ {}^{\circ
}}}$ and $\mu_{Rb} \simeq 47{{ {}^{\circ }}}$) is small and
no clear distinction between KC$_{60}$ and RbC$_{60}$ can be
identified at this point. In contrast the intermolecular
C$_{60}$ environments are different. There is only one type
of environment in KC$_{60}$ (along $\left[ 111\right]$ and
$\left[1\overline{1}1\right]$) where a 'double' bond
approximately face a pentagon from the neighboring C$_{60}$,
as shown in fig.3(a). In RbC$_{60}$ the lower space group
symmetry implies that the $\left[111\right]$ and
$\left[1\overline{1}1\right]$ intermolecular environments
are different, as shown in fig.3(b) and (c). It is possible
that alkali-C$_{60}$ interactions favor roughly the same
chain orientation angle in KC$_{60}$ and RbC$_{60}$ ($\sim
45-50{{}^\circ}$), while the C$_{60}$-C$_{60}$ interactions
determine the relative chain orientations and thus the space
group symmetry. Preliminary calculations\cite{ulter} have
indeed shown that the C$_{60}$-C$_{60}$ intermolecular
potential varies appreciably with the chain orientations,
the symmetry of their arrangement and the unit cell
parameters. This should be kept in mind when analyzing
pressure effects and pressure induced transitions in
AC$_{60}$ compounds\cite {auban,hone,simovic}.

MAS-NMR spectra of RbC$_{60}$ and CsC$_{60}$ are very
similar and  differ from that of
KC$_{60}$\cite{alloul,rachdi}. Alloul et al.\cite{alloul}
suggested that the distribution of spin density along a
chain is influenced by  its neighbors. With the present
results, we can now attribute  the difference between the
KC$_{60}$ and RbC$_{60}$ spectra to the distinct  relative
chain orientations in the two compounds (fig.1(d) and (e)).
The similarity of the RbC$_{60}$ and CsC$_{60}$ spectra
suggests that the chain orientations are likely the same in
RbC$_{60}$ and CsC$_{60}$. As discussed in the introduction,
the physical properties of RbC$_{60}$ and CsC$_{60}$ are
very similar and differ markedly from those of KC$_{60}$. A
strong correlation  between physical properties and relative
chain orientations can thus be  inferred in  polymerized
AC$_{60}$. Electronic structure calculations have already
been performed by Erwin et al.\cite{erwin} and by Tanaka et
al.\cite{tanaka}  for I2/m RbC$_{60}$. Further theoretical
investigations taking  into account the distinct chain
orientations in AC$_{60}$ are much awaited.

\acknowledgments

J.H. and A.Z. acknowledge support from the U.S. Department
of Energy under Contract No. DE-AC03-76SF00098.


\begin{references}
\bibitem{absorb} The DW values may be slightly
underestimated since the data were not corrected for
absorption.
\bibitem{comment}  In principle the monoclinic distortion
should lead to more than 6 orientational variants. However
the distortion is not observed within the resolution of the
present study, and its effect can be safely neglected. We
also mention that we have assumed equivalent domains with
chain orientations $\mu $ and $-\mu $  to calculate the
structure factors $F_{calc}$.
\end{references}
\end{document}